\documentclass[twocolumn,showpacs,preprintnumbers,amsmath,amssymb,prb,aps]{revtex4-1}
\usepackage{epsfig}
\usepackage{graphicx}
\usepackage{dcolumn}
\usepackage{bm}
\usepackage{textcomp}
\usepackage{subfig}

\begin{document}

\title{Fabrication of Simple Apparatus for Resistivity Measurement in High Temperature Range 300-620 K}
\author{Saurabh Singh}
\altaffiliation{Electronic mail: saurabhsingh950@gmail.com}
\author{Sudhir K. Pandey}
\affiliation{School of Engineering, Indian Institute of
Technology Mandi, Kamand - 175005, India}

\date{\today}
\begin{abstract}
A simple and low cost apparatus has been designed and built to measure the electrical resistivity, ($\rho$), of metal and semiconductors in 300-620 K temperature range. The present design is suitable to do measurement on rectangular bar sample by using conventional four-probe dc method. A small heater is made on the sample mounting copper block to achieve the desired temperature. Heat loss from sample holder is minimize by using very low thermal conductive insulator block. This unique design of heater and minimized heat loss from sample platform provide uniform sample temperature and also have very good thermal stability during the measurement. The electrical contacts of current leads and potential probes on the sample are done by using very thin (42 SWG) copper wires and high temperature silver paste. The use of limited components and small heater design make present instrument very simple, light weight, easy to sample mount, small in size, and low cost. To calibrate the instrument pure nickel sample was used, and two other materials La$_{0.7}$Sr$_{0.3}$MnO$_{3}$ (LSMO) and LaCoO$_{3}$ (LCO) were also characterized to demonstrate the accuracy of this set-up. $\rho$(T) behavior on these samples were found to be in good agreement with the reported data. The metal-insulator transition for LSMO (T$_{MI}$ = $\sim$358 K) and the insulator-metal transition for LCO (T$_{IM}$ = $\sim$540 K) were clearly observed and these transitions temperature were also consistent with those reported in literature. 

\end{abstract}

\maketitle
\section{Introduction} 
The world-wide interest of new materials for alternate source of energy and other advanced engineering applications in electronic industry, has led to the need for inspection tools capable of measuring unusual electrical anomalies in various type of materials.\cite{Snyder, Rowe, DMrowe} In the past several decades, various physical properties such as thermoelectric effect, electronic phase transitions (insulator to metal transitions or vice versa), charge ordering, etc have been observed in varieties of materials like transition metal oxides, half-Heusler alloys, clathrates, skutterdites, etc.\cite{Imada, Tritt, Sales, Nolas} To characterize these materials and in the search of new materials, characterization of the electrical resistivity ($\rho$) parameter among the electronic transport coefficients have a significant importance as it is one of the most sensitive indicators of effect driven by the change in electronic band structure. Also, in search of thermoelectric (TE) materials as a renewable source of energy, the characterization of TE material is incomplete without measurement of resistivity as this is one of the crucial parameter in the expression of power factor defined as $\alpha$$^{2}$/$\rho$, where $\alpha$ and $\rho$ is the Seebeck coefficient and electrical resistivity, respectively. Thus, an accurate measurement of $\rho$(T) in wide temperature range are very important for fundamental understanding of phase transformations and their kinetics, nature of ground state, electrical, electronic and magnetic properties observed in the various materials (metals, semiconductors, and insulators) and also useful in selection of suitable materials for industrial applications.\\
The temperature dependent resistivity of a material can be calculated from its measured values of resistance. For the resistance measurement, various methods and models have been suggested in the literature and they are classified by the type of sample ( i.e. thin film, single crystal, powder pellet or small crystalline) and geometry of the contact. The different methods used so far are van der Pauw, Montgometry, Smits, two probes, four probes, etc.\cite{Pauw, Smits, Montgomery} In these methods, Montgomery, van der Pauw and Smith techniques can be used for pellets and bulky samples, whereas two probes can be used for higher resistive (semiconductor) samples and four probes method for low resistive and single crystals.\\
All across the world, many experimental devices have also been fabricated in the laboratory, which can measure the $\rho$(T) in wide temperature region.\cite{Abadlia, Ponnambalam, Ravichandran, Paul, Burkov, Zhou, Guan, Kolb, Michael, Kallaher, Boor, Rawat} The motive behind fabrications in most of these set-up were to study the thermoelectric properties of the materials as they have the option to measure the figure-of-merit, \textit{ZT} = [$\alpha$$^{2}$\textit{T}/($\rho$$\kappa$)], or at least power factor, PF = $\alpha$$^{2}$/$\rho$, where $\alpha$, $\rho$, $\kappa$ and \textit{T} are the Seebeck coefficients, electrical resistivity, thermal conductivity, and absolute temperature, respectively. In most of these setups where options of other electronic transport coefficients along with $\rho$(T) are available there multiple heaters (main furnace along with gradient heater necessary for $\alpha$ measurement), numbers of temperature sensor, temperature controllers for controlling the temperature at two or more points, spring loading arrangements, etc have been used. Although, they provide the facilities of simultaneous measurement of multiple transport coefficients in one set-up, but including all these arrangement results the system complex, bulky and costly.\\
These devices are useful to do the resistivity measurement as long as the study of thermoelectric properties of the materials are concerned, where error of 10\% or more than that in measurement of $\rho$(T) (thus, around same error in calculation of \textit{ZT}). This amount of errors in \textit{ZT} are generally acceptable across the world-wide research community because of the complexity in characterization of transport parameters in high temperature region. In many cases, accurate estimation of physical parameters are required where information of electronic structure are used for the quantitative understanding of the physical property. In general, electrical resistivity of metal and semiconductor systems depends on carrier densities, carrier mobility, relaxation time of the charge carriers. Therefore, to get precise information of activation energy of semiconductor, carrier densities, relaxation time of charge carriers for metal, semiconductor system an accurate measurements of $\rho$(T) become essential.\cite{Mott} Thus, main objective of this paper is to design and fabricate a simple and low cost setup which can measure the high temperature resistivity as accurate as possible.\\

In this paper, we report design and fabrication of simple set-up used for the temperature dependent resistivity measurement. The low cost and commonly available materials are used for the various components of instrument. We employ four probe method and measurement are carried out in 300-620 K temperature range under the vacuum atmosphere. Almost, all the parts can be easily replaced in case of any damage. The sample exchange is very sample as the design of sample holder platform is very simple. A gypsum rectangular block of very low thermal conductivity is attached at one end of heater and sample platform, which minimize the heat loss and provide thermally stable condition during measurement. The information regarding various components, method of calibration, and design are described in detail. For the calibration of set-up, pure nickel sample is used. Two other test materials, La$_{0.7}$Sr$_{0.3}$MnO$_{3}$ and LaCoO$_{3}$, having the electronic phase transition in between 300-620 K temperature range were also characterized. In order to see the precision, accuracy and reproducibility of data, multiple time measurements were performed on each samples. The resistivity data obtained by designed set-up were found to be in good agreement with reported data. The metal-insulator (T$_{\textit{MI}}$ transition at $\sim$358 K for La$_{0.7}$Sr$_{0.3}$MnO$_{3}$ sample and insulator-metal (T$_{\textit{IM}}$ at $\sim$540 K for LaCoO$_{3}$ sample were observed. For the both, La$_{0.7}$Sr$_{0.3}$MnO$_{3}$ and LaCoO$_{3}$ sample, temperature associated with electronic phase transitions were consistent with earlier report.     

\section{Apparatus description}
  The schematic diagram of the resistivity measurement set-up in 300-620 K temperature range is shown in Fig. 1, where the various components are represented by numbers. The selection of various components in the present design are in such a way that the possible challenges of high temperature fabrication have been well taken care. From the mechanical, thermal, electrical and vacuum point of view, the suitable and very low cost materials are considered. A copper plate, \textbf{\textit{10}}, having the length of $\sim$50 mm and thickness of $\sim$2 mm is used for both holding the sample , \textbf{\textit{9}}, and making the heater, \textit{\textbf{7}}, to reach the desired sample temperature. To raise the sample temperature a copper platform is chosen due to its high thermal conductivity ($\kappa$$_{Cu}$ $\sim$4.01 W cm$^{-1}$ K$^{-1}$ at 300 K and $\sim$3.79 W cm$^{-1}$ K$^{-1}$ at 600 K), these values are almost comparable to $\kappa$ of Silver and gold.\cite{Ho} The purpose of selecting copper material over silver and gold is also due to its easy availability and relatively very low cost. A size of suitable dimension ($\sim$50 mm length, $\sim$5 mm width $\&$ $\sim$2 mm thickness) of one end of copper plate, \textbf{\textit{10}}, is used to make the heater. At this end, flexible mica sheet of good thermal conductive and electrical insulating property is wrapped and over this a kanthal wire of 40 SWG ($\sim$0.12 mm) diameter, which resistance is $\sim$35 mm, is tightly wound in $\sim$15 mm in length to make the resistive heater. The use of mica sheet in between kanthal wire and copper plate avoid the direct electrical contact with each other, which also protect the heater from any electrical short circuit. High temperature epoxy cement is applied over the kanthal wire to safe the heater, as at high temperature recoiling and breaking issues in thin wire occurs. Each end of the kanthal wire is joined with copper wire, \textbf{\textit{13}}, for supplying the current in heater. The design of such a small heater closed to sample position have small exposure to vacuum environment, which also minimize the heat loss.  
\begin{figure}[htbp]
 \vspace{0.0cm}
  \begin{center}
  
   \includegraphics[width=0.45\textwidth]{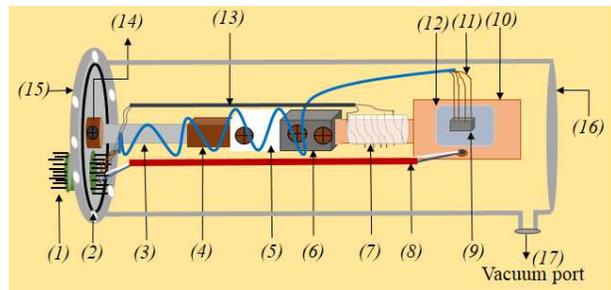}
    \label{Schematic diagram of instrument}
    \captionsetup{justification=raggedright,
singlelinecheck=false
}
    \caption{(Color online) Schematic representation of the instrumental arrangement for the measurement of electrical resistivity in the temperature range 300–620 K.}
    \vspace{0.0cm}
  \end{center}
\end{figure}
  
In the high temperature resistivity measurement heat loss is a major factor in the error contribution. The significant temperature gradient created across the sample due to heat loss give rise to thermo-emf voltage which also gets added in to the measurement of actual sample voltage. To minimize the conductive heat loss from sample supporting copper plate, \textit{\textbf{10}}, a rectangular gypsum bar, \textbf{\textit{6}}, is attached at the heater end side using screw. Thermal conductivity of the gypsum is very low ($\sim$0.17 W/mK at room temperature), and also effective for thermal insulation in high temperature region. The dimensions of used gypsum block are $\sim$55 mm in length, $\sim$15 mm in width, and $\sim$10 mm in thickness. For the mechanical support and stability, gypsum block are connected back to back with rectangular teflon bar, \textit{\textbf{5}} (45 mm $\times$ 15 mm $\times$ 10 mm), brass rectangular bar, \textit{\textbf{4}} (30 mm $\times$ 12 mm $\times$ 10 mm), and stainless steel rod, \textit{\textbf{3}}( $\sim$110 mm length and $\sim$6 mm diameter). The ss rod were further attached at the centre of Stainless steel flange, \textit{\textbf{15}}, using the circular brass disc, \textit{\textbf{14}} (30 mm diameter and $\sim$8 mm thickness). circular o-ring, \textbf{\textit{2}}, is attached to the ss flange, 15, which helps to seal the vacuum chamber. A small groove of rectangular shape (10 mm $\times$ 10 mm) and depth of $\sim$0.5 mm is made on the sample platform. In that groove very thin mica sheet, \textit{\textbf{12}} ($\sim$0.05 mm thickness), which is electrically insulator but a good thermal conductor, is fixed using the high temperature silver paste. The sample is mount on this mica sheet, and electrical insulation is obtained between sample and copper block which acts as a sample supporting platform. To measure the sample temperature, very thin K-type thermocouple, \textbf{\textit{8}}, of 32 SWG ($\sim$0.27 mm) is inserted in a drilled hole near the sample and fixed with using high temperature paste. For current supply to the sample and droped voltage measurement, four copper wires having coating of electrical insulation, \textbf{\textit{11}} (42 SWG i.e. $\sim$0.1 mm diameter), were used. The overall length of sample holder is $\sim$27 cm. The whole sample holder attached to ss flange, 15, was inserted in a cylindrical vacuum chamber, \textbf{\textit{16}}. The inner diameter, outer diameter and length of the vacuum chamber are $\sim$32 cm, $\sim$110 mm, and $\sim$116 mm, respectively. A vacuum port, \textbf{\textit{17}}, of KF-25 size is made at the bottom of the vacuum chamber, which is used to connect the vacuum pump. A rotary pump is used to create the vacuum atmosphere and rotary level ($\approx$ 10$^{-3}$ mbar) of vacuum is maintained during the measurement. This level of vacuum condition is helpful to avoid the sample from oxidation and also minimize the convective heat loss in high temperature measurements.   

\section{Measurement procedure}
A standard four probe method is employed for $\rho$(T) measurement, and schematic representation of this method is shown in Fig. 2. For a sample in rectangular shape, four copper wires are attached linearly on the sample surface using high temperature silver paste. Generally, rectangular shape of sample is used for the resistivity measurement, and typical dimension of common laboratory sample have size of length (L) = 5-10 mm, width (w) = 2-4 mm, and thickness (t) = 1-2 mm range. A Keithley sourcemeter (model 2604 B) is connected to outer leads, \textbf{A} \& \textbf{D}, to provide the constant current I. The potential drop is measured by Keithley nanovoltmeter (2182A) connected to two inner leads \textbf{B} \& \textbf{C}. For the metal and semiconductor resistivity measurement, typical current values range from 1 to 100 mA. At applied constant current I$_{1}$ to the sample the measured potential drop is V$_{1}$, where V$_{1}$ = V$_{IR}$ + V$_{TEP}$. The first term V$_{IR}$ is the resistive voltage at I$_{1}$ and V$_{TEP}$ is the thermal emf contribution. To nullify the thermo-emf present in the circuit, current (I$_{2}$) is immediately reversed (negative) to measure V$_{2}$ where V$_{2}$ = -V$_{IR}$ + V$_{TEP}$. The reversal of current and averaging are necessary to remove the thermal voltage as V$_{TEP}$ can be quite large in thermoelectric materials. The sample resistance Rs,
\begin{equation}
  R_{s} = \frac{V_{1}-V_{2}}{2I_{av}} = \frac{(V_{IR} + V_{TEP})-(-V_{IR}+V_{TEP})}{I_{1}-I_{2}} = \frac{V_{IR}}{I_{av}}  
\end{equation}
In the present case, current I$_{2}$ is equal in magnitude to I$_{1}$ but negative in the sign. From the measured value of resistance R$_{s}$, the resistivity is estimated using the following expression,
\begin{equation}
  \rho = \frac{R_{s}A}{l_{v}}
\end{equation}
where A is the cross sectional area of the sample, i.e. w$\times$t, and l$_{v}$ is the distance between the middle two voltage leads, \textbf{B} and \textbf{C}.
\begin{figure}[htbp]
 \vspace{0.0cm}
  \begin{center}
  
   \includegraphics[width=0.35\textwidth]{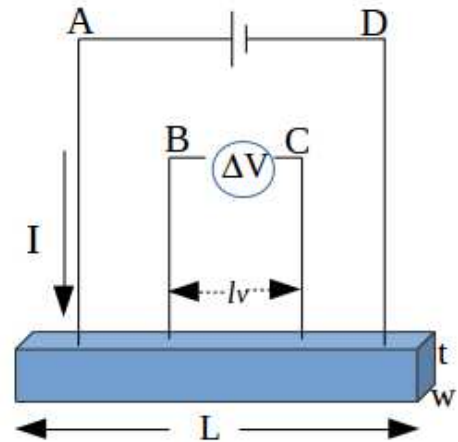}
    \label{Ni resistivity plot}
    \captionsetup{justification=raggedright,
singlelinecheck=false
}
    \caption{(Color online) Schematic representation of principle of electrical resistivity measurement using four probe method.}
    \vspace{0.0cm}
  \end{center}
\end{figure}

 The sample is mount on the sample platform using silver paste. At the bottom surface of the sample, a small amount of silver paste is applied at two outer ends \textbf{A} and\textbf{ D} in such a way that conducting silver paste is not enter in between the applied current region. Therefore, any silver paste out side of the \textbf{A} and \textbf{D} region do not have any affect in the resistivity measurement. After the contacts of voltage and current leads are made on the sample, it is left to dry the silver paste for 6 to 7 hr. To perform the $\rho$ measurement, current is supplied to the kanthal heater, \textbf{\textit{7}}, by using the Crown made dual output dc regulated power supply (0-30V/ 5A). Temperature of the sample is measured by K-type thermocouple, connected with Fluke 17B digital multimeter. The Keithley source meter 2604B and nanovoltmeter 2182 are connected to the computer using GPIB cable, and these two measuring instruments are controlled by the Labview program. To obtained the data at a particular temperature, program is written in such a way that a constant current is passed through the sample and voltage is measured, and immediately current is reversed and again the voltage drop is measured. This process is repeated 10 times at the same temperature and averaged value of sample resistance is measured. The dimensions of the sample is provided in the program to calculate the corresponding values of resistivity. The measurement is allowed in the thermally stable condition such that data is acquired within the change of 0.1 $^{0}$C for a particular temperature. 
\section{Results and Discussion}
In order to validate the fabricated instrument, three different sample i.e. pure nickel (purity $\geq$99.99 \textdiscount, purchased from Sigma-Aldrich), La$_{0.7}$Sr$_{0.3}$MnO$_{3}$, and LaCoO$_{3}$, were chosen. The electrical resistivity measurements on these three samples were carried out using four probe method in the temperature range 300-620 K. Each sample was taken in the rectangular bar shape. In the further discussions of the $\rho$(T), the name of nickel, La$_{0.7}$Sr$_{0.3}$MnO$3$, and LaCoO$_{3}$ samples are coded as Ni, LSMO, and LCO, respectively. Now, the detailed observation of resistivity data will be discussed one by one.

\subsection{Measurement on standard sample (pure nickel)}
 In order to calibrate the designed set-up, resistivity measurement on the high purity metal sample is more appropriate from the reliability and precision point of view. A rectangular nickel sample was taken as a reference materials for characterization purpose. The dimensions of Ni sample used for $\rho$(T) measurement were $\sim$6 mm in length, $\sim$4 mm in width, and $\sim$1.3 mm in thickness. The $\rho$ vs T plot in 300-620 K range is shown in Fig. 3. The whole measurement on same sample were done three times for checking the data reproducibility of the instrument. Fig. 3a show the plots of three time measured data and plotted $\rho$(T) were named as Trial-1, Trial-2, and Trial-3. The averaged of theses three plots have also been added in the same figure. For the comparison purpose, averaged plot are represented in Fig. 3b together with the resistivity data of nickel sample measured by other authors.\cite{Haynes, Abadlia, Ponnambalam, Burkov} In the entire temperature range, $\rho$(T) behavior are in good agreement with those of literature. The values of $\rho$ at 300 K and 620 K are $\sim$7.40 $\mu$$\Omega$ cm and $\sim$28.68 $\mu$$\Omega$ cm, respectively. The maximum value of standard deviation in resistivity data is found to be $\sim$0.15 $\mu$$\Omega$ cm at 500 K, which is $\sim$0.8 \% of the $\rho$ (18.91 $\mu$$\Omega$ cm) value at this temperature. From the averaged data of three trial measurement, the maximum deviation of Trial-1, Trial-2, and Trial-3 are $\sim$0.67 $\mu$$\Omega$ cm (at 500 K), $\sim$0.13 $\mu$$\Omega$ cm (at 350 K), and 0.72 $\mu$$\Omega$ cm (at 500 K) are noticed, and these values are $\sim$3.5\%, $\sim$1.3\%, and $\sim$3.8\%, of their respective observed value of $\rho$. Our data is in excellent agreement with Abadlia \textit{et al}. As shown in Fig. 3b, the maximum deviation between our data and Abadlia \textit{et al} is $\sim$0.35 $\mu$$\Omega$ cm (at $\sim$580 K), which is $\sim$1.4 \% of the observed value of $\rho$.
\begin{figure}[htbp]
 \vspace{0.7cm}
  \begin{center}
  
   \includegraphics[width=0.28\textwidth]{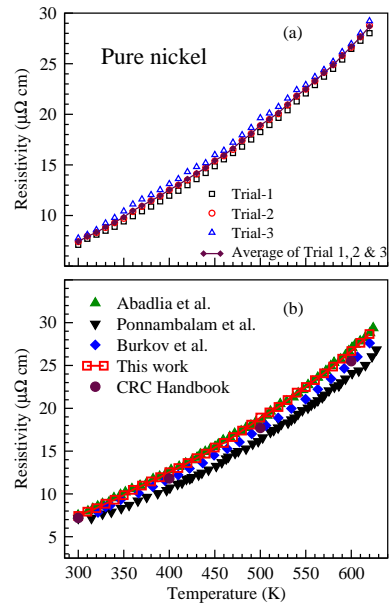}
    \label{resistivity plot}
    \captionsetup{justification=raggedright,
singlelinecheck=false
}
    \caption{(Color online) (a) Temperature dependent resistivity of Pure nickel for three different trial. (b) Our data (this work) compared to that of Abadlia [Ref. \textbf{12}], Ponnambalam [Ref. \textbf{13}], Burkov [Ref. \textbf{16}], and CRC Handbook values [Ref. \textbf{11}].}
    \vspace{0.0cm}
  \end{center}
\end{figure}

\subsection{La$_{0.7}$Sr$_{0.3}$MnO$_{3}$ (LSMO)}
Fig 4. shows the temperature dependent resistivity data obtained for LSMO. In order to check the data reproducible quality of instrument, the measurement have been performed two times on same sample and plots are named as Trial-1 and Trial-2. For the Trial-1, the observed value of $\rho$ at 300 K is $\sim$0.04 $\Omega$ cm and above this temperature an increasing trend in the values of $\rho$ are noticed up to $\sim$358 K. At 358 K, the value of $\rho$ is $\sim$0.057 $\Omega$ cm. For the temperature above $\sim$358 K, continuous decreasing trend is noticed in the values of $\rho$ up to 620 K. At 620 K, the value of $\rho$ is found to be $\sim$3 $\times$10$^{-3}$ $\Omega$ cm. For this system, a metal to insulator transition is reported around 358 K in the literature.\cite{salazar} In order to probe this electronic phase transition more clearly, the measurement of $\rho$ was taken at 1 K temperature interval in 354-365 K range. The data obtained from our set-up also shows metal to insulator transition at $\sim$358 K. To confirm this transition temperature in our sample and to check the accuracy of the designed set-up, measurement was repeated and obtained data is shown by Trial-2 plot in Fig. 4. In Trial-2 measurement, the transition temperature is noticed $\sim$357 K. These T$_{IM}$ values are also consistent with the critical temperature, T$_{C}$, observed around 360 K for the LSMO.\cite{Subhrangsu, Ziese, Lofland} For the sake of clarity, the region around transition temperature is shown in the inset of Fig. 4. The maximum deviation between Trial-1 and Trial-2 measurement is of $\sim$2.36 $\times$ 10$^{-3}$ $\Omega$ cm, which is $\sim$5 \% of the observed $\rho$ data.
\begin{figure}[htbp]
 \vspace{0.7cm}
  \begin{center}
  
   \includegraphics[width=0.40\textwidth]{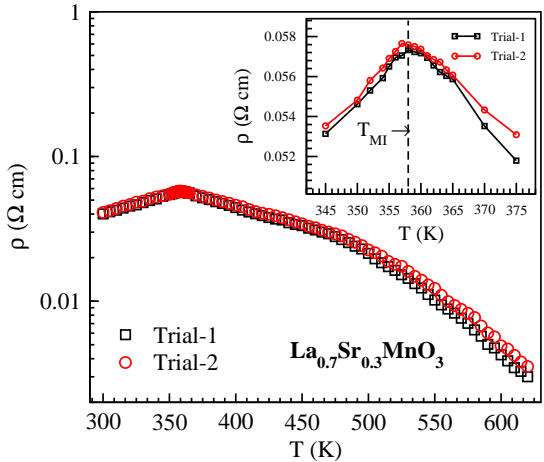}
    \label{LSMO resistivity plot}
    \captionsetup{justification=raggedright,
singlelinecheck=false
}
    \caption{(Color online) The temperature dependent resistivity of La$_{0.7}$Sr$_{0.3}$MnO$_{3}$.}
    \vspace{0.0cm}
  \end{center}
\end{figure} 

\subsection{LaCoO$_{3}$ (LCO)}
Fig 5. shows $\rho$(T) data for LCO in the temperature range 300-620 K. For this system, $\rho$(T) measurement were also performed two times. As shown in Fig. 5, data obtained from our set up are named as Trial-1 and Trial-2. For Trial-1 measurement, the value of $\rho$ at 300 K is $\sim$4.18 $\times$ 10$^{4}$ m$\Omega$ cm. The decrease in magnitude of $\rho$ are noticed in temperature range 300-620 K. The rate of change in $\rho$ is very large in the 300-540 K range, whereas a small change is observed in 540-620 K temperature range. The temperature dependent behavior of $\rho$(T) obtained on our sample is similar to those reported in the literature. The values of $\rho$ are found to be $\sim$4.79 $\times$ 10$^{4}$ m$\Omega$ cm and $\sim$1.95 $\times$ 10$^{4}$ m$\Omega$ cm at 540 K and 620 K, respectively. The large change in values of $\rho$ in 300-540 K range and a small change of $\sim$3 $\times$ 10$^{4}$ m$\Omega$ cm in 540-620 K range indicate an insulator to metal transition, T$_{IM}$, at 540 K.
\begin{figure}[htbp]
 \vspace{0.5cm}
  \begin{center}
  
   \includegraphics[width=0.40\textwidth]{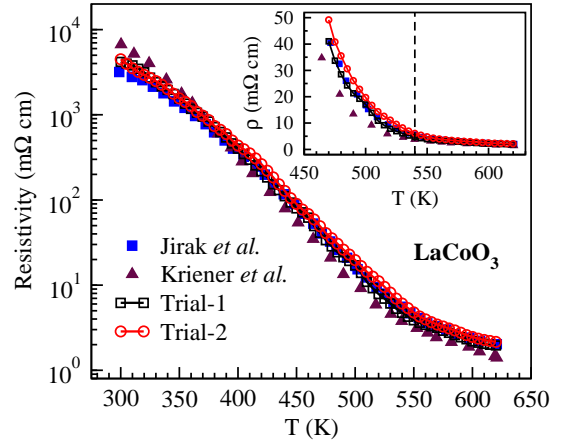}
    \label{LaCoO3 resistivity}
    \captionsetup{justification=raggedright,
singlelinecheck=false
}
    \caption{(Color online) Temperature dependent resistivity of LaCoO$_{3}$ for two different trial. Our data (Trial-1 and trial-2) compared to that of Jirak [Ref. \textbf{32}] and Kriener [Ref. \textbf{33}].}
    \vspace{0.0cm}
  \end{center}
\end{figure} 
For the comparison purpose, resistivity data obtained by \textit{Jirak} \textit{et al} and \textit{Kriener et al} are also plotted in Fig. 5.\cite{Jirak, Kriener} The region around insulator to metal transition is shown in the inset of Fig. 5 for the sake of clarity. This transition temperature, T$_{IM}$ = 540 K, is also consistent with our earlier work where Seebeck coefficient ($\alpha$) measurement was performed on the same sample, i.e. LCO1000, in 300-600 K temperature range.\cite{saurabh}\\
For the comparison purpose, the literature data plotted in the Fig. 3b, 4, and 5 are extracted from the literature figure using the data digitization technique. It is important to notice that digitization process always have an inherent error in data selection from the literature graph. Thus, some deviation ($\sim$0.5 $\mu$$\Omega$ cm for Ni data) are expected due to contribution of digitization error. For LSMO and LCO sample, the difference in magnitude of $\rho$(T) from the reported values can be possible as in case of semiconductor sample, magnitude of resistivity varies from sample to sample due to difference in synthesis conditions. The electronic phase transition temperature for both the samples, LSMO and LCO, is in good agreement with the reported value shows that designed set-up is more versatile for the resistivity characterization of various type of systems.
\section{Conclusions}
In conclusion, a simple and low cost apparatus is fabricated for high temperature resistivity measurement in 300-620 K range. Pure nickel sample is used for the calibration of the instrument. The temperature dependent behavior as well as values of $\rho$(T) obtained from our set-up were found to be in very good agreement with the literature data. The standard deviation in the $\rho$(T) measurement for nickel sample were within the $\pm$1.5 \%. The good quality of data was obtained by achieving the uniform and thermally stable temperature across the sample, which was possible due to design of small heater close to the sample mounting position and minimization of heat loss using thermally insulating gypsum block. The designed set-up is validated by performing the $\rho$(T) measurement on two other samples, La$_{0.7}$Sr$_{0.3}$MnO$_{3}$ and LaCoO$_{3}$, which electronic phase transition temperature are different from each other. The metal-insulator transition temperature for La$_{0.7}$Sr$_{0.3}$MnO$_{3}$ and insulator to metal transition temperature for LaCoO$_{3}$ sample were found to be $\sim$358 K and $\sim$540 K, respectively. The transition temperatures observed in both compounds are consistent with the reported temperature in literature. The use of small heater, simple design, and limited components of low cost materials make the present setup very small in size and light weight. The accuracy in measurement of $\rho$(T) values and its capability of probing the electronic phase transitions shows that fabricated setup can be very helpful in understanding the electronic structure as well as characterization of the various type of materials for many industrial applications.  
\section{ACKNOWLEDGMENTS}
We sincerely thank R. S. Raghav and other workshop staff for their technical support in the fabrication process of the vacuum chamber and sample holder parts.


\begin{thebibliography}{99}

\bibitem{Snyder} G. J. Snyder and E. S. Toberer, Nat. Mater. \textbf{7}, 105 (2008).

\bibitem{Rowe} D. M. Rowe, \textit{CRC Handbook of Thermoelectrics} (CRC Press, Boca Raton, 1995).

\bibitem{DMrowe} D. M. Rowe, \textit{Thermoelectrics Handbook: Macro to Nano}, edited by D. M. Rowe (CRC, Boca Raton, 2006).

\bibitem{Imada}M. Imada, A. Fujimori, and Y. Tokura, \textit{Rev. Mod. Phys.}, \textbf{70}, 4 (1998).
\bibitem{Tritt} \textit{Harvesting Energy Through Thermoelectrics: Power Generation and Cooling}, edited by T. M. Tritt and M. A. Subramanian, \textit{MRS Bulletin Theme on Thermoelectrics}, Vol. \textbf{31}, (2006).
\bibitem{Sales}B. C. Sales, D. Mandrus, and R. K. williams, \textit{Science} \textbf{272}, 1325 (1996).
\bibitem{Nolas} G. S. Nolas, J. L. Cohn, G. A. Slack, and S. B. Schujman, \textit{Appl. Phys. Lett.} \textbf{73}, 178 (1998).
\bibitem{Pauw} L. J. van der Pauw, \textit{Philips res. Rep.} \textbf{13}, 1 (1958).
\bibitem{Smits} F. M. Smits, \textit{Bell Syst. Tech. J.} \textbf{37}, 711 (1958).
\bibitem{Montgomery} H. C. Montgomery, \textit{J. Appl. Phys.} \textbf{42}, 2971 (1971).
\bibitem{Haynes} W. H. Haynes, \textit{CRC Handbook of Chemistry and Physics}, 94th ed. (CRC Press, 2013).
\bibitem{Abadlia} L. Abadlia, F. Gasser, K. Khalouk, M. Mayoufi, and J. G. Gasser, \textit{Rev. Sci. Instrum.} \textbf{85}, 095121 (2014).
\bibitem{Ponnambalam}V. Ponnambalam, S. Lindsey, N. S. Hickman, and T. M. Tritt, \textit{Rev. Sci. Instrum.} \textbf{77}, 073904 (2006).
\bibitem{Ravichandran}J. Ravichandran, J. T. Kardel, M. L. Scullin, J. H. Bahk, H. Heijmerikx, J. E. Bowers, and A. Majumdar, \textit{Rev. Sci. Instrum.} \textbf{82}, 015108 (2011).
\bibitem{Paul}B. Paul, \textit{Measurement} \textbf{45}, 133 (2012).
\bibitem{Burkov}A. T. Burkov, A. Heinrich, P. P. Konstantinov, T. Nakama, and K. Yagasaki, \textit{Meas. Sci. Technol.} \textbf{12}, 264 (2001).
\bibitem{Zhou}Z. Zhou and C. Uher, \textit{Rev. Sci. Instrum.} \textbf{76}, 023901 (2005).
\bibitem{Guan}A. Guan, H. Wang, H. Jin, W. Chu, Y. Guo, and G. Lu, \textit{Rev. Sci. Instrum.} \textbf{84}, 043903 (2013).
\bibitem{Kolb}H. Kolb, T. Dasgupta, K. Zabrocki, E. Mueller, and J. de Boor, \textit{Rev. Sci. Instrum.} \textbf{86}, 073901 (2015).
\bibitem{Michael} P. H. Michael Bottger, E. Flage-Larsen, O. B. Karlsen, and Terje G. Finstad, \textit{Rev. Sci. Instrum.} \textbf{83}, 025101 (2012).
\bibitem{Kallaher} R. L. Kallaher, C. A. Latham, and F. Sharifi,\textit{ Rev. Sci. Instrum.} \textbf{84}, 013907 (2013).
\bibitem{Boor}J. de Boor, C. Stiewe, p. Ziolkowski, T. Dasgupta, G. Karpinski, E. Lenz, F. Edler, and E. mueller, \textit{J. Elect. Mater.}, \textbf{42} 7 (2013).
\bibitem{Rawat} P. K. Rawat and Biplab Paul, \textit{Measurement}, \textbf{91}, 613 (2016).
\bibitem{Mott} N.F. Mott, E.A. Davis, \textit{Electronic Processes in Non Crystalline Materials}, Clarendon Press, Oxford, 1971.
\bibitem{Ho} C. Y. Ho, R. W. Powell, and P. E. Liley, \textit{J. Phys. Chem.} Ref. Data, \textbf{1}, 279 (1972).
\bibitem{Manzello} S. L. Manzello, S. Park, T. Mizukami, and D. P. Bentz, “Measurement of thermal properties of gypsum board at elevated temperatures,” in \textit{Proceedings of the Fifth International Conference on Structures in Fire} (SiF08) (Organising Committee, Fifth International Conference Structures in Fire (SiF’08), 2008), pp. 656-665.
\bibitem{Heikes} R. R. Heikes, R. C. Miller, and R. Mazelsky, \textit{Physica} \textbf{30}, 1600 (1964).
\bibitem{salazar} D. Salazar, D. Arias, O.J. Dura , M.A. Lopez de la Torre, \textit{Journal of Alloys and Compounds}, \textbf{583}, 141 (2014).
\bibitem{Subhrangsu} S. Taran, B. K. Chaudhuri, S. Chatterjee, H. D. Yang, S. Neeleshwar and Y. Y. Chen, \textit{J. Appl. Phys.} \textbf{98}, 103903 (2005).
\bibitem{Ziese}M. Ziese, \textit{J. Phys.: Condens. Matter}, \textbf{13}, 2919 (2001).
\bibitem{Lofland} S. E. Lofland, V. Ray, P. H. Kim, S. M. Bhagat, M. A. Manheimer, and S. D. Tyagi, \textit{Phys. Rev. B}, \textbf{55}, 2749 (1997).
\bibitem{Jirak}Z. Jirak, J. Hejtmanek, K. Knizek, and M. Veverka, \textit{Phys. Rev. B}, \textbf{78}, 014432 (2008).
\bibitem{Kriener} M. Kriener, C. Zobel, A. Reichl, J. Baier, M. Cwik, K. Berggold, H. Kierspel, O. Zabara, A. Freimuth, and T. Lorenz, \textit{Phys. Rev. B} \textbf{69}, 094417 (2004).
\bibitem{saurabh} Saurabh Singh and Sudhir K. Pandey, \textit{Measurement}, \textbf{102}, 26 (2017).

\end{thebibliography}
\end{document}